\documentclass[twoside,11pt]{article}
\usepackage[colorlinks,bookmarksopen,bookmarksnumbered,citecolor=blue,urlcolor=red]{hyperref}
\usepackage{setspace}
\usepackage[margin=1in,right=1in,top=1in,bottom=1in]{geometry} 
\usepackage{amsmath,amsthm,amssymb,scrextend, enumerate, mathrsfs}
\usepackage{graphicx}
\usepackage{fancyhdr}
\usepackage{tikz}
\usepackage{pgfplots}
\usepackage{natbib}
\usepackage{amsmath, threeparttable}
\usepackage[labelfont=bf]{caption}
\usepackage[mathscr]{euscript}
\usepackage[scr]{rsfso}
\usepackage{amsthm}
\usepackage{amssymb}
\usepackage{multicol}
\usepackage{multirow}
\usepackage{booktabs}
\usepackage{lscape}

\renewcommand{\subsectionmark}[1]{}

\begin{document}

\title{A General Framework for Designing and Evaluating Active-Controlled Trials with Non-Inferiority Objectives}

\author{Antonio Olivas-Martinez\textsuperscript{1}, Fei Gao\textsuperscript{2}, and Holly Janes\textsuperscript{2} \\\\
\textsuperscript{1}Department of Biostatistics, University of Washington\\
\textsuperscript{2}Bioinformatics and Epidemiology Program Vaccine and Infectious Disease Division,\\ Fred Hutch Cancer Center
}

\maketitle

\begin{abstract}
Active-controlled trials with non-inferiority objectives are often used when effective interventions are available, but new options may offer advantages or meet public health needs. In these trials, participants are randomized to an experimental intervention or an active control. The traditional non-inferiority criterion requires that the new intervention preserve a substantial proportion of the active control effect. A key challenge is the absence of a placebo arm, which necessitates reliance on historical data to estimate the active control effect and assumptions about how well this effect applies to the target population. Another challenge arises when the active control is highly effective, as the new intervention may still be valuable even if it does not meet the traditional criterion. This has motivated alternative criteria based on sufficient efficacy relative to a hypothetical placebo. In this work, we propose a general framework for designing and evaluating non-inferiority trials that integrates all existing analytical methods and accommodates both traditional and alternative success criteria. The framework enables the systematic comparison of methods in terms of type I error, power, and robustness to misspecification of the active control effect. We illustrate its applicability in the design of a future HIV prevention trial with a highly effective active control. In this application, our framework identifies methods that provide greater efficiency and robustness than commonly used approaches and demonstrates practical advantages of the alternative non-inferiority criterion. Overall, this framework offers a comprehensive toolkit for rigorous non-inferiority trial design, supporting method selection and the evaluation of new interventions.
    
\end{abstract}

\section{Introduction}

Active-controlled trials are the most frequently used phase 2b/3 trial design when a current intervention is known to be effective and is available, but new interventions are still warranted \citep{rothmann2011design, fleming2011some, food2016non}. The new intervention may have superior or clinically-relevant efficacy and advantages in terms of reduced cost or toxicity, improved tolerability, adherence, or ease of implementation; or additional products may be needed to serve public health needs. In such contexts, a placebo-controlled design may not be justified. In active-controlled trials, participants are randomized to one or more experimental interventions or to an active control, without a placebo arm. They may be designed to assess whether the new intervention is superior or non-inferior to the active control, where being non-inferior commonly is defined as preserving a fraction of the active control effect, often $50\%$, a standard known as the preservation of effect criterion \citep{ food2016non,fleming2008current}. Active-control trials designed to assess non-inferiority are often called non-inferiority trials.

A significant challenge in drawing inferences from active-controlled trials is the absence of a placebo arm. Assessing non-inferiority traditionally involves estimating the active control effect relative to placebo based on historical data, preferably from randomized trials, necessitating assumptions about the extent to which the historical estimate of the active control effect applies to the target population. These assumptions must take into account possible effect modification by both measured and unmeasured modifiers, as well as factors such as advancements in concomitant care, shifts in disease etiology or diagnostic criteria, evolving trial endpoints, and changes in the dose or regimen of the active control as its clinical use evolves \citep{fleming2011some}. One commonly invoked assumption in this setting is constancy, which essentially posits that the active control effect estimated in the historical trial(s) remains unchanged in the target population. Regrettably, any deviations from this assumption may undermine the validity of conclusions drawn from non-inferiority trials \citep{ rothmann2011design, fleming2011some,food2016non,wang2002utility,odem2013adjusting}.

The primary methods for assessing non-inferiority based on active-controlled trials are the fixed-margin and synthesis methods \citep{rothmann2011design, food2016non,wang2002utility}. The fixed-margin method utilizes a predetermined margin to evaluate the efficacy of the experimental intervention against the active control, while the synthesis method integrates data from both the active-controlled trial and historical trial(s) in order to assess non-inferiority. Both methods have been extensively evaluated under constancy and non-constancy conditions. While the traditional synthesis method only controls type-I error under constancy, the fixed-margin method is demonstrated to be robust to some deviations from constancy \citep{wang2002utility,odem2013adjusting,james2003some, hung2007issues, snapinn2008controlling, brittain2012valid}. A generalization of the synthesis method improves robustness to non-constancy \citep{odem2013adjusting}.

Another challenge with non-inferiority design arises in the context of a highly effective active control.  In such a context, an experimental intervention may still have public health impact even if it would not satisfy commonly employed preservation of effect non-inferiority criteria. For example, if the active control has 95$\%$ prevention efficacy, a typical 50$\%$ preservation of effect criterion would stipulate that a new intervention have a prevention efficacy greater than 77.6$\%$, corresponding to a 50\% preservation of the log hazard ratio. Such a high bar may not be appropriate if the new intervention has advantages in terms of individual preference, cost, or feasibility of implementation.  A compelling success criterion may be that the new intervention meets the success criterion that would have been used had a placebo-controlled design been possible.  This is the spirit of the inferred efficacy criterion proposed for supporting the design of COVID-19 vaccine non-inferiority trials \citep{fleming2021covid}. A unified approach for comparing the operating characteristics of the various analytical methods for different non-inferiority criteria is lacking.

In light of the aforementioned challenges, we propose a general framework for evaluating the non-inferiority of experimental interventions compared to active controls. Our framework accommodates both traditional and new non-inferiority criteria and all existing methods for non-inferiority analysis, and facilitates the systematic evaluation and comparison across methods based on operating characteristics that are either defined conditional on the historical data or unconditional.  It also enables quantification of the degree of non-constancy accommodated by each method, and offers a design approach that formally factors in uncertainty in the active control effect based on historical data. We demonstrate the applicability of our framework by designing a future HIV prevention trial with a highly effective active control.

\section{Motivating Example}\label{sec:hiv_prev_motivation}

Much success has been seen in recent years in biomedical prevention of HIV, with oral \citep{grant2010preexposure, baeten2012antiretroviral, thigpen2012antiretroviral, van2012preexposure, marrazzo2015tenofovir} and now injectable \citep{landovitz2021cabotegravir, delany2022cabotegravir, bekker2024twice, purpose2} antiretrovirals proven effective for prevention, known as pre-exposure prophylaxis (PrEP).  Long-acting Cabotegravir (CAB-LA) has proven highly effective and superior to oral PrEP in both women and men who have sex with men \citep{landovitz2021cabotegravir, delany2022cabotegravir}. However, rollout, uptake, and adherence remain significant challenges, and additional interventions are clearly needed to achieve the Joint United Nations Programme on HIV/AIDS (UNAIDS) targets for HIV incidence \citep{joint2023path, van2014perspectives, henderson2023future}, including an effective HIV vaccine. HIV monoclonal antibodies, on-demand products, and alternative antiretrovirals and delivery devices are under investigation \citep{corey2021two, walker2021amp}. In a setting where CAB-LA is licensed and available for use, a potential trial design to evaluate a novel antiretroviral as PrEP could be an active-controlled trial with CAB-LA as the active control, which seeks to establish that the novel antiretroviral is non-inferior to CAB-LA.

A key target population for such a trial would be women in sub-Saharan Africa, who remain at high risk of HIV despite advances in HIV prevention \citep{joint2023path,murewanhema2022hiv,moyo2022long}. In this population, the CAB-LA prevention efficacy (PE), as measured by one minus the hazard ratio, has been recently estimated at $92.8\%$ with a $95\%$ confidence interval (CI): $76.1\%– 97.8\%$ \citep{donnell2023counterfactual}. 
Applying the conventional 50\% preservation of effect criterion would require a new intervention to exceed 73.2\% PE, a threshold that could eliminate from consideration promising interventions.  Details of this calculation are provided in Section~\ref{sec:ni_h0}. This motivates the use of an alternative criterion, assessing success relative to the threshold that would have applied in prior placebo-controlled PrEP trials, which corresponded to a minimum PE of 30\% \citep{grant2010preexposure,baeten2012antiretroviral,van2012preexposure, marrazzo2015tenofovir}. A new antiretroviral that achieves at least $30\%$ PE and offers advantages such as lower cost, improved adherence, reduced toxicity, or greater ease of implementation could have substantial public health impact. This example highlights the need for a flexible framework that accommodates such alternative non-inferiority criteria while maintaining rigorous statistical evaluation.

\section{Non-Inferiority Hypothesis and Identification Challenges}\label{sec:ni_h0}

We begin by introducing notation and framing the null hypothesis within our statistical framework. For an intervention of interest $A$, let $\psi_A$ denote a function characterizing the distribution of an outcome of interest in a population receiving intervention $A$. We define the effect of intervention $A$ relative to intervention $B$ as $\gamma_{AB}:=\psi_A-\psi_B$. This formulation accommodates various types of outcomes, including binary, continuous, count, and censored event-time outcomes.

For example, in HIV prevention trials, the outcome of interest is often time to HIV acquisition, a censored event-time outcome. In such settings, the effect of intervention $A$ relative to $B$ is commonly summarized using the log hazard ratio (log HR), either assumed constant over time or evaluated at a fixed landmark time. Let $h_A(t)$ and $h_B(t)$ denote the hazard functions under interventions $A$ and $B$, respectively, where $h_A(t)$ represents the instantaneous risk of HIV acquisition at time $t$ for individuals receiving intervention $A$. At a fixed time $t_0$, e.g., two years, the effect contrast $\gamma_{AB}$, i.e., the log hazard ratio $\log\{h_A(t_0)/h_B(t_0)\}$, can be written as $\psi_A - \psi_B$ where $\psi_A=\log\{h_A(t_0)\}$ and $\psi_B=\log\{h_B(t_0)\}$. While our framework focuses on effect measures expressible as $\gamma_{AB}=\psi_A - \psi_B$, prevention efficacy (PE) is also widely reported in HIV prevention, typically defined as $1 - \text{HR} = 1 - \exp(\gamma_{AB})$. Although PE is not itself in the additive form we target, it is a monotone transformation of the log HR and is therefore implicitly included in analyses based on this contrast.

We denote the placebo, active control, and experimental intervention as $P$, $C$, and $X$ respectively. Then, $\gamma_{XP}$, $\gamma_{CP}$, and $\gamma_{XC}$ are their relative effects in the target population,  i.e., in the population from which the active-controlled trial is sampled. The effect of the experimental intervention relative to placebo can be decomposed as $\gamma_{XP}=\gamma_{XC}+\gamma_{CP}$. When negative values of the effect indicate a benefit, the scientific null hypothesis is:
\begin{equation}
H_0:\quad\gamma_{XC}+\gamma_{CP}\ge \Delta,\label{h0}
\end{equation}
where $\Delta$ denotes the null efficacy that must be ruled out. For the traditional preservation of effect non-inferiority criterion, $\Delta = f\gamma_{CP}$, with $f\in(0,1)$ denoting the fraction of the active control effect to be preserved. For the novel inferred efficacy criterion, $\Delta=\Delta_0$, where $\Delta_0$ is a fixed minimally acceptable level of efficacy. It generalizes the concept of demonstrating superiority relative to a hypothetical placebo, which has been considered in previous literature \citep{james2003some, snapinn2008controlling, snapinn2004alternatives}, and in this case sets $\Delta_0 = 0$. In public health settings where a minimum prevention efficacy is often required---such as in our HIV prevention motivating example---the threshold $\Delta_0$ is set below zero (recall that negative values indicate a benefit) to reflect this minimally acceptable level of efficacy rather than simple superiority.

The scientific null hypothesis~\eqref{h0} hinges on $\gamma_{CP}$, a parameter unidentifiable from the active-controlled trial alone. However, historical placebo-controlled trials provide evidence as to the active control effect in the historical setting. Let $\gamma_{CP,H}$ denote the active control's effect in the historical setting, reflecting its performance in the population from which the historical trials were drawn, administered following those trials' protocols, including adherence and relevant modifiers. The traditional synthesis method assumes $\gamma_{CP}=\gamma_{CP,H}$, a premise known as the constancy assumption \citep{food2016non, fleming2008current,wang2002utility, james2003some}. The fixed-margin method assumes $\gamma_{CP} = \gamma_{CP,H} + a$, where $a \geq 0$ reflects a conservative adjustment based on the precision of the historical estimate and tends to zero as that estimate becomes more precise. Importantly, this adjustment is defined in the context of a conditional null hypothesis that depends on the observed historical data, whereas the null hypothesis we consider in this work is unconditional and accounts for uncertainty in the historical estimate. The specific form of $a$ and this distinction will be detailed in Section~\ref{sec:special_cases}. A standard assumption is also that normally distributed and consistent estimators $\hat\gamma_{XC}$ for $\gamma_{XC}$ and $\gamma_{CP,H}$ for $\hat\gamma_{CP,H}$ exist with variances $V_{XC}$ and $V_{CP,H}$, and that $\hat\gamma_{XC}$ and $\hat\gamma_{CP,H}$ are independent. While the assumptions regarding the estimators are justifiable, being based on independent data from large-scale trials, the validity of the constancy assumption is questionable, and requires careful scrutiny.

In the context of HIV prevention, relevant effect modifiers may include participants’ age, sex, and sexual behavior, which could influence the efficacy of the active control intervention. In our motivating example, a reasonable estimate for the effect of CAB-LA among sub-Saharan African women is provided by \cite{donnell2023counterfactual}, who used data from three contemporary studies conducted across five countries (Botswana, Kenya, Malawi, South Africa, and Zimbabwe). Their analysis adjusted for potential effect modifiers such as age and baseline diagnosis of sexually transmitted infections, specifically gonorrhea or chlamydia, to account for differences in sexual behavior. They estimated a log HR of $\hat\gamma_{CP,H} = \log(0.072) \approx -2.64$, corresponding to a 92.8\% prevention efficacy. If one assumes constancy, ie., assumes $\gamma_{CP} = \gamma_{CP,H}$, and defines non-inferiority using a 50\% preservation of effect definition, an intervention must have a 73.2\% PE or higher to be non-inferior to CAB-LA ($\Delta = -1.32$). On the other hand, if non-inferiority is defined using the 30\% inferred efficacy criterion ($\Delta = \log(0.7) \approx -0.36$), meaning that an intervention must have 30\% PE or higher to be non-inferior to CAB-LA, this corresponds to preservation of at least 13.6\% of the CAB-LA effect.

\section{The General Framework}

This section introduces our general framework for evaluating non-inferiority methods. We begin by defining key parameters that allow us to translate a range of existing methods into this unified structure. The interpretations of these parameters are made relative to the scientific null hypothesis of interest considered in this framework.

To quantify the degree of non-constancy that a given method can accommodate, we parameterize deviations from constancy. Let $\lambda_0:=\left(\gamma_{CP}-\gamma_{CP,H}\right)/\gamma_{CP,H}$ denote the \textit{true relative effect deviation}. Here, $\lambda_0=0$ corresponds to constancy. Assuming that negative effect values indicate benefit, $\lambda_0<0$ means that the active control is less effective in the target population than in the historical population, while $\lambda_0>0$ indicates greater effectiveness. This parameterization yields $\gamma_{CP} = (1+\lambda_0)\gamma_{CP,H}$, allowing the scientific null hypothesis~\eqref{h0} to be written as
\[
H_0:\quad \gamma_{XC}+(1-f)(1+\lambda_0)\gamma_{CP,H}\ge\Delta_0,
\]
where $\Delta_0=0$ for the preservation of effect criterion and $f=0$ for the inferred efficacy criterion.

This formulation of the null hypothesis replaces $\gamma_{CP}$ with $\lambda_0$.  Both $\gamma_{CP}$ and $\lambda_0$ are unidentifiable. The utility of this formulation of the scientific null hypothesis is to show how it relates to what we call the \textit{operational null hypothesis} which is actually tested using commonly-employed non-inferiority methods.  Specifically, a value for $\lambda_0$ is assumed, which we denote by $\lambda_1$. For example, one may assume that $\lambda_1 = 0$, corresponding to constancy, or assign another value, which could be chosen based on historical data. Using $\lambda_1$, we define the operational null hypothesis as:
\begin{equation}
H_0^{op}:\quad \gamma_{XC}+(1-f)(1+\lambda_1)\gamma_{CP,H}\ge\Delta_0\label{h0_op}.
\end{equation}
This is the hypothesis that is actually tested using data from both the active-controlled trial and historical trials, as opposed to the scientific null hypothesis~\eqref{h0}, which relies on unidentifiable parameters. Distinguishing between scientific and operational null hypotheses is important for comparing type-I error rates under non-constancy across methods, as these rates are evaluated at the boundary of the scientific null hypothesis.

Although one might object that we are simply replacing $\lambda_0$ with $\lambda_1$, working with $\lambda_1$ offers practical advantages. It provides an interpretable way to characterize departures from constancy across different settings and clarifies how methods incorporate historical data. Our framework makes these assumptions explicit and place them on a common scale. This will become clearer in the next subsection.

Within this general framework, we test the operational null hypothesis~\eqref{h0_op} using the following test statistic:
\begin{equation}
T_{u,\lambda_1,f,\Delta_0} =\frac{\hat\gamma_{XC}+(1-f)(1+\lambda_1)\hat\gamma_{CP,H}-\Delta_0}{\sqrt{V_{XC}+u^2(1-f)^2(1+\lambda_1)^2V_{CP,H}}}.\label{t_um}
\end{equation}
Here, $u\ge0$ is the \textit{unifying} parameter that determines how uncertainty from the historical estimate $\hat\gamma_{CP,H}$ is incorporated when combining it with the active-controlled trial data. Setting $u=0$ recovers fixed-margin methods, which treat the historical effect as fixed and exclude its variability, whereas $u=1$ recovers synthesis methods, which propagate the full variance of the historical estimate. Values $u>1$ further inflate the contribution of $V_{CP,H}$, yielding more conservative tests (see Section~\ref{sec:special_cases}). Thus, $u$ provides a single tuning mechanism for how different approaches weight historical uncertainty. Because the denominator of~\eqref{t_um} grows with $u$, increasing $u$ reduces the magnitude of the test statistic under both null and alternative hypotheses. This enhances robustness of type I error control under misspecification of the historical active-control effect, but at the expense of reduced power. These trade-offs will be quantified in Section~\ref{sec:op_characteristics}. As previously discussed, $\lambda_1$ encodes our assumption about the relative effect deviation $\lambda_0$, while $f$ and $\Delta_0$ specify the success criterion. In particular, $\Delta_0=0$ and $f\in(0,1)$ yields the preservation of effect criterion, whereas $f=0$ and $\Delta_0<0$ corresponds to the new inferred efficacy criterion.

In Appendix~\ref{sec:appA_formulas}, we show that at the boundary of the operational null hypothesis, i.e., when $\gamma_{XC}+(1-f)(1+\lambda_1)\gamma_{CP,H}=\Delta_0$, the test statistic $T_{u,\lambda_1,f,\Delta_0}$ is normally distributed with mean zero and variance
\[
\sigma^2=\frac{V_{XC}+(1-f)^2\widetilde{V}_{CP,H}}{V_{XC}+u^2(1-f)^2(1+\lambda_1)^2V_{CP,H}},
\]
where
\[
\widetilde{V}_{CP,H}:=\begin{cases}
    (1+\lambda_1)^2V_{CP,H},& \text{if }u>0,  \\[6pt]
    V_{CP,H}, & \text{if } u=0 \text{ (fixed-margin methods).} 
\end{cases}
\]
The quantity $\widetilde{V}_{CP,H}$ unifies the representation of historical variability across methods: it reduces to $V_{CP,H}$ under fixed-margin methods ($u=0$) and to $(1+\lambda_1)^2V_{CP,H}$ otherwise; see details in Appendix~\ref{sec:appA_formulas}. We reject the operational null hypothesis when $T_{u,\lambda_1,f,\Delta_0}$ falls below $-Z_{1-\alpha}$, where $Z_{1-\alpha}$ denotes the $(1-\alpha)$ quantile of the standard normal distribution.

\subsection{Special Cases within Our Framework}\label{sec:special_cases}

We now show that our general framework accommodates most existing methods for assessing non-inferiority (see Table~\ref{tab:l0min}).

\begin{table}[]
\caption{Common methods for assessing non-inferiority and their robustness to type I error control under constancy}
  \resizebox{\textwidth}{!}{
\begin{tabular}{@{}lccccc@{}}
\toprule
Method                               & \multicolumn{2}{c}{Parameter}                                                                                                                                     & Is unconditional type I error $\le\alpha$ under constancy?$\dag$
 \\ \midrule
                                     & Unifying ($u$)                           & Assumed relative effect deviation ($\lambda_1$)&                                                                                                                               &                      \\ 
\midrule
Traditional synthesis method$^\text{(a)}$      & 1                       & 0                                                                                                                               & Yes                \\
Bias-adjusted synthesis method$^\text{(b)}$                 & 1                       & $\lambda_1$                                                                                                         & Yes$^*$    \\
Odem-Davis method$^\text{(b)}$                 & $(1+\lambda_1)^{-1}$ & $\lambda_1$ & Yes$^*$   \\  \midrule
Fixed-margin 95-95 method$^\text{(a)}$           &  0                       & $1.96\sqrt{V_{CP,H}}/\hat\gamma_{CP,H}$           &                                                                                                             Yes    \\
Fixed-margin 0-95 method$^\text{(c)}$                           & 0                       & 0           &                                                                                                             No  \\ \midrule
\end{tabular}
}
\footnotesize{
$\dag\alpha$ denotes the nominal level. For the 95-95 and 0-95 methods, $\alpha=0.025$. $^*$This condition is satisfied when $\lambda_1<0$, which is common for these methods. (a) \cite{food2016non,fleming2008current,wang2002utility}; (b) \cite{odem2013adjusting}; (c) \cite{wang2002utility}.  }\label{tab:l0min}
\end{table}

\subsubsection{Synthesis Methods}

Synthesis methods represent a simple strategy to incorporating historical information. These methods directly combine data from the historical and active-control trials, and importantly, their operational null hypothesis aligns exactly with that of~\eqref{h0_op} \citep{rothmann2011design, food2016non,fleming2008current,wang2002utility,odem2013adjusting}.

The so-called \textit{bias-adjusted synthesis method} assumes a relative effect deviation of $\lambda_1$ and directly combines $\hat\gamma_{XC}$ and $\hat\gamma_{CP,H}$ to construct an estimator of $\gamma_{XC}+(1-f)(1+\lambda_1)\gamma_{CP,H}$, with corresponding variance $V_{XC}+(1-f)^2(1+\lambda_1)^2V_{CP,H}$. The operational null hypothesis~(\ref{h0_op}) is rejected if the upper bound of the $(1-2\alpha)\%$ CI for this quantity lies below $\Delta_0$. Specifically, this method rejects (\ref{h0_op}) when
\[
\hat\gamma_{XC}+(1-f)(1+\lambda_1)\hat\gamma_{CP,H} + Z_{1-\alpha}\sqrt{V_{XC}+(1-f)^2(1+\lambda_1)^2V_{CP,H}}<\Delta_0,
\]
which is equivalent to the following test statistic
\[
T_{SM,\lambda_1} = \frac{\hat\gamma_{XC}+(1-f)(1+\lambda_1)\hat\gamma_{CP,H}-\Delta_0}{\sqrt{V_{XC}+(1-f)^2(1+\lambda_1)^2V_{CP,H}}}
\]
being less than $-Z_{1-\alpha}$. This corresponds to our general test statistic in~\eqref{t_um} with parameters $(u,\lambda_1)=(1,\lambda_1)$. When constancy is assumed---i.e., when $\lambda_1=0$---we refer to this as the \textit{traditional synthesis method}.

Given that synthesis methods lack robustness to deviations beyond the assumed relative effect deviation $\lambda_1$---particularly in their control of type I error (as will be clarified in the next section)---\cite{odem2013adjusting} proposed a modification that avoids contracting the variance of $(1-f)(1+\lambda_1)\hat\gamma_{CP,H}$ by the factor $(1+\lambda_1)^2$. This variant uses the same point estimator for $\gamma_{XC}+(1-f)(1+\lambda_1)\gamma_{CP,H}$, but assumes a larger variance $V_{XC}+(1-f)^2V_{CP,H}$. The resulting test statistic is
\[
T_{OD,\lambda_1} = \frac{\hat\gamma_{XC}+(1-f)(1+\lambda_1)\hat\gamma_{CP,H}-\Delta_0}{\sqrt{V_{XC}+(1-f)^2V_{CP,H}}},
\]
which corresponds to the general form~\eqref{t_um} with parameters $(u,\lambda_1)=(\{1+\lambda_1\}^{-1},\lambda_1)$. Notably, this method is only applicable and meaningful when $\lambda_1 \neq 0$. We refer to it as the \textit{Odem-Davis method}.

\subsubsection{Fixed Margin Methods}

Fixed-margin methods are the most widely used methods for incorporating historical information in non-inferiority testing \citep{rothmann2011design, food2016non,fleming2008current,wang2002utility}. Unlike synthesis methods, a fixed-margin method treats the historical data as fixed and known, and assumes that $\gamma_{CP}$ can be conservatively approximated by $\tilde\gamma_{CP}:=\hat\gamma_{CP}+a$ where $a=Z_{1-\theta}\sqrt{V_{CP,H}}$ and $\theta\in(0,0.5]$, typically $\theta=\alpha$. The term $a$ represents the absolute deviation from the point estimate $\hat\gamma_{CP,H}$ used to construct the upper bound of a $100(1 - 2\theta)\%$ CI for $\gamma_{CP,H}$. This serves as a conservative adjustment to account for uncertainty in the historical estimate. A fixed success margin is then defined as $\delta:=\Delta_0-(1-f)\tilde\gamma_{CP}$. The method tests the following operational null hypothesis:
\begin{equation}
H_0^\text{FM}: \quad \gamma_{XC} \ge \delta.\label{h0_fm}
\end{equation}
Assuming $\gamma_{CP,H}=\hat\gamma_{CP,H}$ and setting $\lambda_1=Z_{1-\theta}\sqrt{V_{CP.H}}/\hat\gamma_{CP,H}$, the operational null hypothesis in~\eqref{h0_fm} is equivalent to that in~\eqref{h0_op}. The fixed-margin method rejects~\eqref{h0_fm} if the upper bound of the $100(1-2\alpha)\%$ CI for $\gamma_{XC}$ falls below $\delta$, i.e, if $\hat\gamma_{XC}+ Z_{1-\alpha}\sqrt{V_{XC}}<\Delta_0-(1-f)(1+\lambda_1)\hat\gamma_{CP,H}$. This is equivalent to the test statistic
\[
T_{FM,\lambda_1} = \frac{\hat\gamma_{XC}+(1-f)(1+\lambda_1)\hat\gamma_{CP,H}-\Delta_0}{\sqrt{V_{XC}}}.
\]
being less than $-Z_{1-\alpha}$. Within our framework, this corresponds to setting $(u,\lambda_1)=(0,Z_{1-\theta}\sqrt{V_{CP.H}}/\hat\gamma_{CP,H})$ in the general statistic (\ref{t_um}). When $\theta=\alpha=0.025$, this yields the well-known \textit{95-95 method}. When $\theta=0.5$ and $\alpha=0.025$, it corresponds to the \textit{0-95 method}, which assumes no effect deviation from the historical estimate.

A subtle but important point arises when representing fixed-margin methods within our framework. To match the operational null hypotheses in~\eqref{h0_op} and~\eqref{h0_fm}, we assume $\gamma_{CP,H}=\hat\gamma_{CP,H}$, which is natural when the historical data are considered fixed and known. In addition, the margin component $\lambda_1\hat\gamma_{CP,H}=Z_{1-\theta}\sqrt{V_{CP,H}}$ is deterministic and does not contribute to uncertainty in the test statistic, which is reflected by setting $u=0$ in~\eqref{t_um}. Thus, in this case, $\lambda_1$ is set to be $Z_{1-\theta}\sqrt{V_{CP.H}}/\hat\gamma_{CP,H}$; in other words, the assumed degree of constancy is proportional to the variability in the historical estimated active control effect.

\subsection{Alternative Conceptual Unification: Snapinn’s Discounting Perspective}

Another attempt to conceptually unify non-inferiority procedures was proposed by \cite{snapinn2004alternatives}, who interpreted fixed-margin methods and preservation of effect criteria as forms of \emph{discounting} historical evidence. In this perspective, historical information is down-weighted or variance-inflated to hedge against the untestable assumptions underlying active-controlled designs, such as constancy. Importantly, from this viewpoint, the preservation of effect criterion is a form of discounting---used to strengthen evidence for superiority relative to a hypothetical placebo---rather than a stand-alone regulatory or scientific objective.

\cite{snapinn2008controlling} formalized this idea using a two-parameter class of test statistics
\[
T^{SJ}_{v,w}=\frac{\hat\gamma_{XC}+(1-w)\hat\gamma_{CP,H}}{\sqrt{V_{XC}+(1-w)^2V_{CP,H}+2v(1-w)\sqrt{V_{XC}V_{CP,H}}}},
\]
where $w\in[0,1]$ is a weighting factor applied to the historical effect, and $v\ge0$ inflates the variance to account for potential deviations from constancy. In this class, the 95–95 fixed-margin method corresponds to $(v,w)=(1,0)$, while synthesis methods correspond to $(v,w)=(0,-\lambda_1)$. Varying $(v,w)$ reproduces many common procedures, enabling evaluation of their operating characteristics under departures from constancy.

Within our general framework, this class is nested as a special case via the mapping
\[
(u,\lambda_1,f,\Delta_0)=\left(\left\{1+\frac{2v\sqrt{V_{XC}}}{(1-w)\sqrt{V_{CP,H}}}\right\}^{1/2}, -w, 0, 0\right),
\]
so that $T^{SJ}_{v,w}$ is recovered from the general test statistic in~\eqref{t_um}. This parameterization clarifies that fixed-margin methods do not “discount” arbitrarily: they correspond to a specific, quantifiable choice of $\lambda_1 = Z_{1-\theta}\sqrt{V_{CP,H}}/\hat\gamma_{CP,H}$ and $u=0$, with $f=0$ and $\Delta_0=0$. Unlike the qualitative discounting lens, our framework makes these assumptions explicit and, as will be demonstrated below, allows evaluation of operating characteristics across true relative effect deviations $\lambda_0$.

Importantly, this formulation distinguishes genuine preservation of effect objectives from discounting: by allowing $f>0$ and $\Delta_0<0$, our framework can represent situations where there is an independent regulatory or scientific interest in preserving a fraction of the active-control effect, rather than solely aiming for superiority relative to a hypothetical placebo.

\section{Operating Characteristics within the General Framework}\label{sec:op_characteristics}

Two types of operating characteristics are pertinent for evaluating non-inferiority in active-controlled trials: conditional and unconditional. Both account for stochasticity in the active-controlled trial, but conditional characteristics assume a fixed and known active control effect in the historical population, while unconditional characteristics treat $\hat\gamma_{CP,H}$ as a random variable with a known variance $V_{CP,H}$. Since unconditional operating characteristics incorporate uncertainty in estimating the active control effect, robustness to constancy assumption violations is most naturally evaluated in terms of unconditional type-I error. We evaluate unconditional operating characteristics and facilitate designing an active-controlled trial to achieve adequate unconditional power while controlling unconditional type-I error.

Under the assumptions outlined in the preceding sections, the unconditional power of the statistic (\ref{t_um}) to reject the scientific null hypothesis (\ref{h0}) when the experimental intervention has an effect size $\gamma_{XP}$ is 
  \begin{equation}
       \Phi\left(\frac{\Delta_0+\left\{(1+\lambda_0)-(1-f)(1+\lambda_1)\right\}\gamma_{CP,H}-\gamma_{XP}-Z_{1-\alpha}\sqrt{V_{XC}+u^2(1-f)^2(1+\lambda_1)^2V_{CP,H}}}{\sqrt{V_{XC}+(1-f)^2\widetilde{V}_{CP,H}}}\right),\label{up_um}
    \end{equation}
   where $\Phi$ is the cumulative distribution function of the standard normal distribution. Its unconditional type-I error at the boundary of the scientific null hypothesis (\ref{h0}) is
\begin{equation}
  \Phi\left(\frac{(1-f)(\lambda_0-\lambda_1)\gamma_{CP,H}-Z_{1-\alpha}\sqrt{V_{XC}+u^2(1-f)^2(1+\lambda_1)^2V_{CP,H}}}{\sqrt{V_{XC}+(1-f)^2\widetilde{V}_{CP,H}}}\right)\label{eq:ut1e_um}.
\end{equation}
Similarly, the conditional power and type-I error are
  \begin{equation}
       \Phi\left(\frac{\Delta_0+\left\{(1+\lambda_0)-(1-f)(1+\lambda_1)\right\}\hat\gamma_{CP,H}-\gamma_{XP}-Z_{1-\alpha}\sqrt{V_{XC}+u^2(1-f)^2(1+\lambda_1)^2V_{CP,H}}}{\sqrt{V_{XC}}}\right)\label{cp_um}
    \end{equation}
    and
    \begin{equation}
  \Phi\left(\frac{(1-f)(\lambda_0-\lambda_1)\hat\gamma_{CP,H}-Z_{1-\alpha}\sqrt{V_{XC}+u^2(1-f)^2(1+\lambda_1)^2V_{CP,H}}}{\sqrt{V_{XC}}}\right)\label{eq:ct1e_um}.
\end{equation}
For completeness, we provide the derivations of expressions~\eqref{up_um}--\eqref{eq:ct1e_um} in Appendix~\ref{sec:appA_formulas}.

To explore the implications of these expressions in practical settings, we focus on a specific region of the parameter space where power is at least 50$\%$ and type-I error is below 50$\%$. In Appendices~\ref{sec:appA_parameter_space} and~\ref{sec:appA_tol_non_constancy}, we show that the parameter space of interest satisfies
\begin{equation}
    \gamma_{XP}<\Delta_0+(1+\lambda_0)\gamma_{CP,H}-(1-f)(1+\lambda_1)\left\{\gamma_{CP,H}+u\:Z_{1-\alpha}\sqrt{V_{CP,H}}\right\}\label{eq:constraint_g_xp}
\end{equation}
and
\begin{equation}
    \lambda_1\le\lambda_0-\frac{(1+\lambda_0)u\:Z_{1-\alpha}\sqrt{V_{CP,H}}}{\gamma_{CP,H}+u\:Z_{1-\alpha}\sqrt{V_{CP,H}}}.\label{eq:constraint_l1}
\end{equation}
We further assume that the active control is efficacious in the historical population, i.e., $\gamma_{CP,H}<0$.

Under these conditions, conditional power always exceeds unconditional power, and conditional type-I error is always lower than unconditional type-I error. This occurs because conditional characteristics ignore the uncertainty associated with estimating the active control effect from historical data. Mathematically, this follows from two key observations: first, expressions~\eqref{up_um} and~\eqref{eq:ut1e_um} reduce to~\eqref{cp_um} and~\eqref{eq:ct1e_um} when $V_{CP,H}=0$ and $\hat\gamma_{CP,H}=\gamma_{CP,H}$; and second, under the constraints~\eqref{eq:constraint_g_xp} and~\eqref{eq:constraint_l1}, the numerators inside the standard normal arguments of~\eqref{up_um} and~\eqref{eq:ut1e_um} are strictly positive and strictly negative, respectively.

Moreover, both power and type-I error decrease as $\lambda_0$ increases. Mathematically, this inverse relationship arises because the numerators inside the standard normal arguments of~\eqref{up_um} and~\eqref{eq:ut1e_um} depend inversely on $\lambda_0$, given that $\gamma_{CP,H}$ is assumed negative. Clinically, this makes sense: an increase from $\lambda_0$ to $\lambda_0'$ implies either (i) less deviation from constancy if $\lambda_0' < 0$, or (ii) that the active control is more effective in the target population than in the historical one if $\lambda_0' \ge 0$. In the first case, the historical estimate becomes more accurate, and in the second, it serves as a more conservative approximation of the active control effect in the target population. In both cases, this leads to reduced type-I error. However, power is also reduced, since the null hypothesis becomes harder to reject when the active control is more effective in the target population.

In the following section, we derive additional insights from expressions~\eqref{up_um}--\eqref{eq:ct1e_um} and introduce further operating characteristics relevant to trial design—both conceptually and through an application of the proposed framework to the design of an HIV prevention trial.

\section{Application to Designing an HIV Prevention Trial}

As introduced in Section~\ref{sec:hiv_prev_motivation}, our goal is to design an HIV prevention trial among sub-Saharan African women using CAB-LA, a highly effective PrEP regimen, as the active control. The outcome of interest is time to HIV acquisition, and the log HR is the measure of effect, which we convert to the prevention efficacy scale for interpretation.

Given the high historical efficacy of CAB-LA (PE of $92.8\%$, $95\%$ CI: $76.1\%– 97.8\%$), we consider both types of non-inferiority success criteria: i) the $50\%$ preservation of effect criterion, parameterized in our framework as $(f,\Delta_0)=(0.5,0)$, and ii) the inferred $30\%$ prevention efficacy criterion, parameterized as $(f,\Delta_0)=(0,\log(0.7))$ on the log HR scale.

We evaluate five analytical methods: the traditional synthesis method, the bias-adjusted synthesis method with $\lambda_1 = -0.23$, the Odem-Davis method with $\lambda_1 = -0.23$, the 95-95 method, and the 0-95 method. The value $\lambda_1=-0.23$ used in the bias-adjusted synthesis and Odem-Davis methods corresponds to a CAB-LA PE of 86.8$\%$, which is one standard deviation below the estimated active control effect on the log HR scale. Using the historical estimates reported by \cite{donnell2023counterfactual} ($\hat\gamma_{CP,H} = \log(0.072) \approx -2.64$ and $\sqrt{V_{CP,H}} = 0.61$), the relative effect deviation implicitly assumed by the 95-95 method is $\lambda_1 = 1.96\sqrt{V_{CP,H}} / \hat\gamma_{CP,H}\approx -0.454$.

\subsection{Maximum Unconditional Power}\label{sec:max_up}

Incorporating the uncertainty in estimating the active control effect from historical data necessarily reduces power. For a design alternative within the restricted parameter space specified in Section~\ref{sec:op_characteristics}, a study can be carried out to achieve any conditional power above 50$\%$. However, there is an upper bound to the corresponding unconditional power, given by
\begin{equation*}
\Phi\left(-uZ_{1-\alpha }+\frac{\Delta_0+\left\{(1+\lambda_0)-(1-f)(1+\lambda_1)\right\}\gamma_{CP,H}-\gamma_{XP}}{(1-f)\sqrt{\widetilde{V}_{CP,H}}}\right).
\end{equation*}
Details of this derivation are provided in Appendix~\ref{sec:appA_parameter_space}. This means that some design alternatives cannot be detected unconditionally with the desired power, affecting the interpretation of findings from active-controlled trials designed to control conditional power.

\begin{figure}[h!]
    \includegraphics[width=1\linewidth]{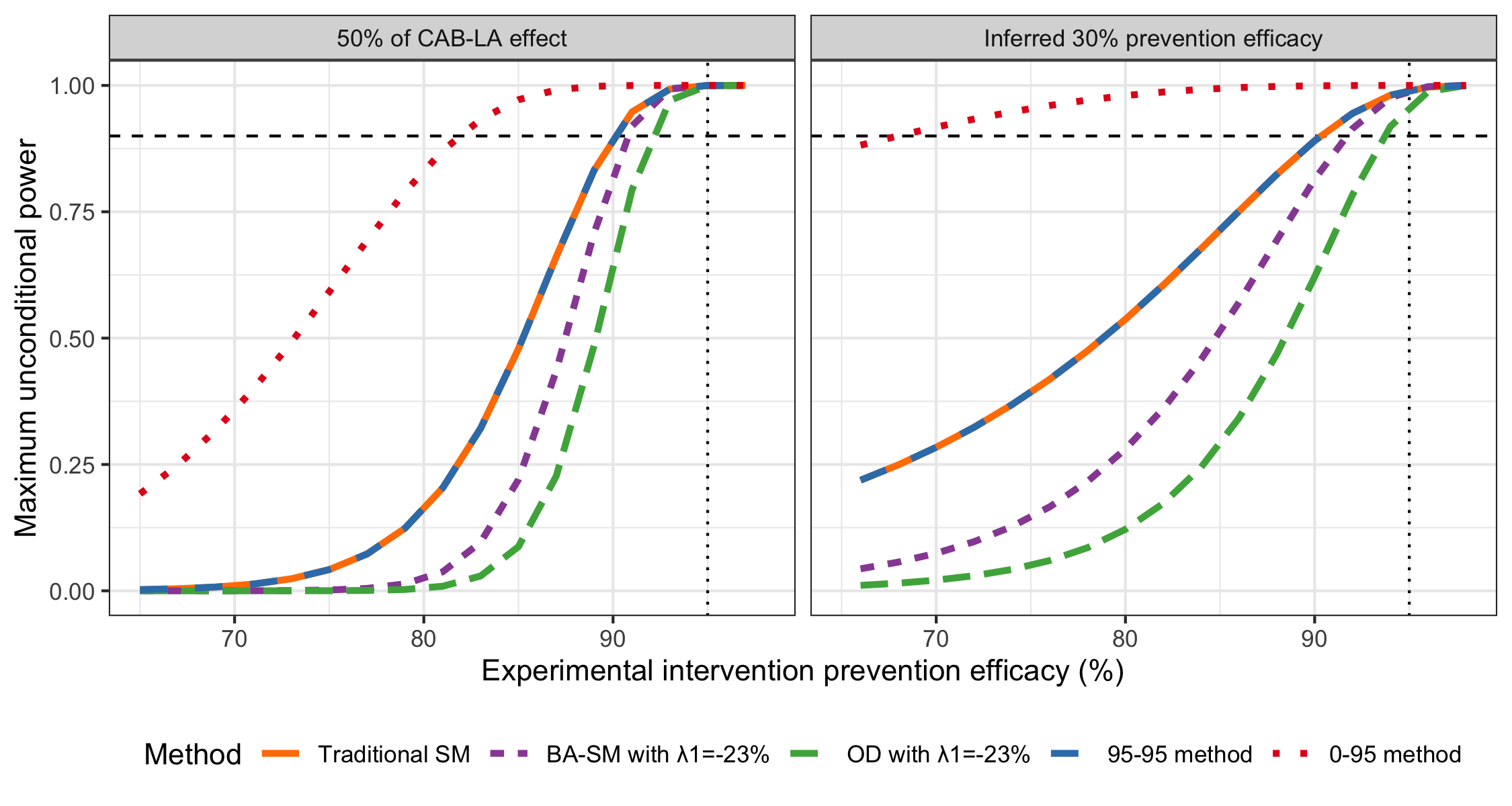}\caption{Maximum unconditional power for detecting design alternatives across our selected methods and criteria. SM, synthesis method; BA-SM, bias-adjusted synthesis method; OD, Odem-Davis method. The horizontal dashed line represents an unconditional power of 0.90. The vertical dotted line represents an experimental intervention with a PE of 95$\%$.}\label{fig:bounds}
\end{figure}

Figure~\ref{fig:bounds} shows the maximum unconditional power for detecting interventions with prevention efficacy between $65\%$ and $98\%$ across the five analytical methods. We see, for example, that a $90\%$ PE intervention cannot be detected with $90\%$ unconditional power with most methods, except with the 0-95 method, while a $95\%$ PE intervention can be detected with $90\%$ unconditional power across all methods and criteria. This underscores the challenges of non-inferiority designs when the active control is highly effective.

\subsection{Trial Design Controlling Unconditional Power}

Active-controlled trials for assessing non-inferiority are typically designed to achieve a target conditional power. However, as previously discussed, conditional power is systematically lower than unconditional power. This discrepancy may result in trials being underpowered, leading to the failure to approve interventions that are in fact non-inferior. We refer to this conventional practice as the \textit{traditional approach} to clinical trial design.

To mitigate potential power loss when the active control proves more effective in the target population than anticipated (i.e., $\lambda_0 > 0$), a common ad hoc strategy is to posit a larger value of $\lambda_0$ at the design stage. By assuming a stronger active control effect than suggested by historical data, this approach inflates both conditional and unconditional power under the planning model. We refer to this strategy as the \textit{ad hoc approach}.

While calculating the sample size to achieve a desired level of conditional power is a well-established procedure (see, e.g., \cite{rothmann2011design}), our framework also enables the design of trials that control unconditional power---the probability of rejecting the scientific null hypothesis when both historical and trial data are treated as random, assuming a fixed deviation from constancy, i.e., a fixed value of $\lambda_0$. We refer to this as the \textit{novel approach}.

Designing a non-inferiority trial to ensure a target level of unconditional power to detect a given design alternative is a multi-step approach. First, determine if the desired power for a design alternative can be achieved through the conditions outlined in Sections~\ref{sec:op_characteristics} and~\ref{sec:max_up}. Second, if achievable, calculate the numerical value for $V_{XC}$ by solving formula~\eqref{up_um} equated to the desired power. Third, determine the sample size $N$ based on the relationship between $V_{XC}$ and sample size in the active-controlled trial. Fourth, evaluate the accommodated level of non-constancy using expression~\eqref{l0min} as described in subsection~\ref{sec:controlled_non_constancy} below.

We apply this procedure to evaluate three trial design strategies: 1) the traditional approach,  targeting $90\%$ conditional power; 2) the novel approach, ensuring $90\%$ unconditional power; and 3) the ad hoc approach, guaranteeing 90\% conditional power under the assumption that CAB-LA has $94.7\%$ PE---an effect half a standard deviation above the estimated active control effect on the log scale. Our design alternative is an experimental intervention with 95\% PE, which we confirmed to be feasible in the previous subsection. The required number of events is computed under constancy, i.e., $\lambda_0 = 0$, and 1:1 allocation. Total sample size is then determined assuming a fixed two-year follow-up period, a 3\% baseline incidence under placebo, 7.5\% annual loss to follow-up, and constant incidence and PE in both groups.

The trial design specifications and operating characteristics for the various approaches, methods, and success criteria considered are summarized in Table~\ref{tab:comp_methods_preserv} for the preservation of effect criterion and in Table~\ref{tab:comp_methods_inf_eff} for the inferred efficacy criterion. Appendix~\ref{sec:r_function} describes the R function we developed for this work and used to generate these tables. To facilitate interpretation, we first introduce two additional operating characteristics within our general framework: the success margin and the tolerable level of non-constancy. We also provide guidance on comparing different success criteria, which is important to consider before reviewing the results presented in the tables.

\begin{table}[h!]
\caption{Design specifications and operating characteristics for detecting a 95\% efficacious intervention with 90\% power under the 50\% effect preservation criterion, by method and design approach.}
\resizebox{\textwidth}{!}{
\begin{tabular}{lcccccc}
\toprule
     \hline
\multicolumn{7}{c}{Traditional approach: Controlling conditional power and assuming CAB-LA PE $=92.8\%$}  \\ \hline \hline
Method & Success margin$^*$ & RNE (total, experimental:control) & Sample size$\dag$ & Controlled non-constancy$\ddag$ & UP0 & UPa \\ 
   \hline
Traditional SM & 3.12 & 19 (8:11) & 5,766 (2,883/arm) & 92.8$\%$ & 0.86 & 0.69 \\ 
  BA-SM, $\lambda_1=-23\%$ & 2.42 & 27 (11:16) & 8,008 (4,004/arm) & 86.8$\%$ & 0.86 & 0.65 \\ 
  OD with $\lambda_1=-23\%$ & 2.21 & 32 (13:19) & 9,510 (4,755/arm) & 84.4$\%$ & 0.86 & 0.63 \\ 
  95-95 method & 2.05 & 37 (15:22) & 11,012 (5,506/arm) & 85.0$\%$ & 0.83 & 0.60  \\ 
  0-95 method & 3.73 & 15 (6:9) & 4,504 (2,252/arm) & 94.8$\%$ & 0.87 & 0.72 \\ \hline\\ \hline\hline 
\multicolumn{7}{c}{Novel approach: Controlling unconditional power and assuming CAB-LA PE $=92.8\%$}  \\ \hline \hline
Method & Success margin$^*$ & RNE (total, experimental:control) & Sample size$\dag$ & Controlled non-constancy$\ddag$ & UP0 & UPa \\ 
   \hline
Traditional SM & 3.07 & 24 (10:14) & 7,208 (3,604/arm) & 92.8$\%$ & 0.90 & 0.75  \\ 
  BA-SM, $\lambda_1=-23\%$ & 2.40 & 33 (14:19) & 10,090 (5,045/arm) & 86.8$\%$ & 0.90 & 0.70 \\ 
  OD with $\lambda_1=-23\%$& 2.17 & 40 (16:24) & 12,012 (6,006/arm) & 84.3$\%$ & 0.90 & 0.69 \\ 
  95-95 method & 2.05 & 52 (21:31) & 15,516 (7,758/arm) & 85.9$\%$ & 0.90 & 0.70 \\ 
0-95 method & 3.73 & 17 (7:10) & 5,046 (2,523/arm) & 94.9$\%$ & 0.90 & 0.77\\ 
   \hline\\
     \hline
     \hline
\multicolumn{7}{c}{Ad hoc approach: Controlling conditional power and assuming CAB-LA PE $=94.7\%$}  \\ \hline \hline
Method & Success margin$^*$ & RNE (total, experimental:control) & Sample size$\dag$ & Controlled non-constancy$\ddag$ & UP0 & UPa \\  \hline
Traditional SM & 2.98 & 33 (16:17) & 11,668 (5,834/arm) & 92.8$\%$ & 0.95 & 0.83 \\ 
  BA-SM, $\lambda_1=-23\%$ & 2.32 & 53 (26:27) & 18,738 (9,369/arm) & 86.8$\%$ & 0.97 & 0.84\\ 
  OD with $\lambda_1=-23\%$ & 2.06 & 71 (35:36) & 25,226 (12,613/arm) & 83.8$\%$ & 0.97 & 0.82 \\ 
  95-95 method & 2.05 & 72 (35:37) & 25,394 (12,697/arm) & 86.8$\%$ & 0.95 & 0.78 \\ 
  0-95 method & 3.73 & 23 (11:12) & 8,236 (4,118/arm) & 95.1$\%$ & 0.95 & 0.85 \\ 
   \hline
   \bottomrule
\end{tabular}
\vspace{0.25em}
}
\footnotesize
SM, synthesis method; BA-SM, bias-adjusted synthesis method; OD, Odem-Davis method; RNE, required number of events assuming 1:1 randomization. The numbers in parentheses correspond to the expected number of events in each group. UP0, unconditional power under constancy (when CAB-LA PE is $92.8\%$); UPa, unconditional power when CAB-LA PE is $94.7\%$. $^*$The experimental intervention is deemed successful if the hazard ratio of HIV acquisition for the experimental intervention relative to the active control is lower than this margin. $\dag$Assuming $3\%$ underlying placebo incidence, a fixed 2-year follow-up, $7.5\%$ loss to follow-up per year, and constant incidence and PE. $\ddag$This is the lowest CAB-LA PE such that the unconditional type-I error is below 0.025. It corresponds to $1-\exp\left([1+\lambda_{0,\min}]\hat\gamma_{CP,H}\right)$.
\label{tab:comp_methods_preserv}
\end{table}

\begin{table}[h!]
\caption{Design specifications and operating characteristics for detecting a 95\% efficacious intervention with 90\% power under the 30\% prevention efficacy criterion, by method and design approach.}
\resizebox{\textwidth}{!}{
\begin{tabular}{lcccccc}
\toprule
     \hline
\multicolumn{7}{c}{Traditional approach: Controlling conditional power and assuming CAB-LA PE $=92.8\%$}  \\ \hline \hline
Method & Success margin$^*$ & RNE (total, experimental:control) & Sample size$\dag$ & Controlled non-constancy$\ddag$ & UP0 & UPa \\ 
   \hline
Traditional SM & 6.13 & 9 (4:5) & 2,882 (1,441/arm) & 92.8$\%$ & 0.83 & 0.73 \\ 
  BA-SM, $\lambda_1=-23\%$ & 3.72 & 15 (6:9) & 4,504 (2,252/arm) & 86.8$\%$ & 0.83 & 0.69 \\ 
  OD with $\lambda_1=-23\%$ & 2.88 & 22 (9:13) & 6,506 (3,253/arm) & 83.7$\%$ & 0.81 & 0.65 \\ 
  95-95 method & 2.94 & 21 (9:12) & 6,486 (3,243/arm) & 87.0$\%$ & 0.78 & 0.63 \\ 
  0-95 method & 9.72 & 7 (3:4) & 2,162 (1,081/arm) & 95.2$\%$ & 0.85 & 0.76 \\ \hline\\ \hline\hline 
\multicolumn{7}{c}{Novel approach: Controlling unconditional power and assuming CAB-LA PE $=92.8\%$}  \\ \hline \hline
  \hline

Method & Success margin$^*$ & RNE (total, experimental:control) & Sample size$\dag$ & Controlled non-constancy$\ddag$ & UP0 & UPa \\ 
   \hline
Traditional SM & 5.67 & 14 (6:8) & 4,324 (2,162/arm) & 92.8$\%$ & 0.90 & 0.81 \\ 
  BA-SM, $\lambda_1=-23\%$ & 3.51 & 24 (10:14) & 7,208 (3,604/arm) & 86.8$\%$ & 0.90 & 0.78 \\ 
  OD with $\lambda_1=-23\%$& 2.50 & 49 (20:29) & 14,514 (7,257/arm) & 83.2$\%$ & 0.90 & 0.76 \\ 
  95-95 method & 2.94 & 48 (20:28) & 14,414 (7,207/arm) & 88.8$\%$ & 0.90 & 0.79 \\ 
0-95 method & 9.72 & 8 (3:5) & 2,502 (1,251/arm) & 95.3$\%$ & 0.90 & 0.83 \\ 
   \hline\\
     \hline
     \hline
\multicolumn{7}{c}{Ad hoc approach: Controlling conditional power and assuming CAB-LA PE $=94.7\%$}  \\ \hline \hline
Method & Success margin$^*$ & RNE (total, experimental:control) & Sample size$\dag$ & Controlled non-constancy$\ddag$ & UP0 & UPa \\  \hline
Traditional SM & 5.72 & 13 (6:7) & 4,804 (2,402/arm) & 92.8$\%$ & 0.89 & 0.81 \\ 
  BA-SM, $\lambda_1=-23\%$ & 3.47 & 25 (12:13) & 8,922 (4,461/arm) & 86.8$\%$ & 0.91 & 0.80 \\ 
  OD with $\lambda_1=-23\%$ & 2.52 & 45 (22:23) & 15,856 (7,928/arm) & 83.2$\%$ & 0.90 & 0.76 \\ 
  95-95 method & 2.94 & 33 (16:17) & 11,668 (5,834/arm) & 88.1$\%$ & 0.86 & 0.74 \\ 
  0-95 method & 9.72 & 8 (4:4) & 2,882 (1,441/arm) & 95.4$\%$ & 0.91 & 0.84 \\ 
   \hline
   \bottomrule
\end{tabular}
\vspace{0.25em}
}
\footnotesize

SM, synthesis method; BA-SM, bias-adjusted synthesis method; OD, Odem-Davis method; RNE, required number of events assuming 1:1 randomization. The numbers in parentheses correspond to the expected number of events in each group. UP0, unconditional power under constancy (when CAB-LA PE is $92.8\%$); UPa, unconditional power when CAB-LA PE is $94.7\%$. $^*$The experimental intervention is deemed successful if the hazard ratio of HIV acquisition for the experimental intervention relative to the active control is lower than this margin. $\dag$Assuming $3\%$ underlying placebo incidence, a fixed 2-year follow-up, $7.5\%$ loss to follow-up per year, and constant incidence and PE. $\ddag$This is the lowest CAB-LA PE such that the unconditional type-I error is below 0.025. It corresponds to $1-\exp\left([1+\lambda_{0,\min}]\hat\gamma_{CP,H}\right)$.
\label{tab:comp_methods_inf_eff}
\end{table}

\subsection{Establishing a Success Margin}

Our framework elucidates that any method within its scope can be regarded as a fixed-margin method. From the definition of our general test statistic~\eqref{t_um}, we find that the rejection region $T_{u,\lambda_1,f,\Delta_0}<-Z_{1-\alpha}$ is equivalent to $\hat\gamma_{XC}+Z_{1-\alpha}\sqrt{V_{XC}}<\delta$, where
\begin{equation}
\delta = \Delta_0-(1-f)(1+\lambda_1)\hat\gamma_{CP,H} - Z_{1-\alpha}\left\{\sqrt{V_{XC}+u^2(1-f)^2(1+\lambda_1)^2V_{CP,H}}-\sqrt{V_{XC}}\right\}.\label{ni_margin}
\end{equation}
 Notably, in general, the margin depends on $V_{XC}$, and thus relies on data from the active-controlled trial. However, for fixed margin methods ($u=0$), the margin simplifies to $\Delta_0-(1-f)(1+\lambda_1)\hat\gamma_{CP,H}=\Delta_0-(1-f)\left(\hat\gamma_{CP,H} + Z_{1-\theta}\sqrt{V_{CP,H}}\right)$, which only depends on historical data and design parameters. Having a defined form for the success margin enables the investigator to characterize at the design stage the size of the experimental to active control effect that would constitute success, for potential values of $V_{XC}$.

The success margins in our HIV prevention trial design example are reported in the second column of Tables~\ref{tab:comp_methods_preserv} and~\ref{tab:comp_methods_inf_eff}, expressed on the relative hazard ratio scale. These margins represent the upper bound for the relative hazard ratio between the experimental and active control interventions, below which the experimental intervention is deemed successful.

\subsection{Quantifying Controlled Non-Constancy}\label{sec:controlled_non_constancy}

Given the need to ensure strict control over type-I error in phase 2b/3 trials and the result that non-constancy can inflate this error, quantifying the level of non-constancy accommodated is crucial. As the true extent of non-constancy remains unknown, our best course of action is to establish thresholds on the maximum degree of non-constancy—or, in simpler terms, the minimum value of $\lambda_0$—where the unconditional type-I error remains at or below $\alpha$. We call this the \textit{tolerable level of non-constancy} which is shown in Appendix~\ref{sec:appA_tol_non_constancy} to be:
\begin{equation}
    \lambda_{0,\min} := \lambda_1+\frac{\sqrt{V_{XC}+u^2(1-f)^2(1+\lambda_1)^2V_{CP,H}}-\sqrt{V_{XC}+(1-f)^2\widetilde{V}_{CP,H}}}{(1-f)\gamma_{CP,H}}Z_{1-\alpha}.\label{l0min}
\end{equation}
The tolerable level of non-constancy allows practitioners to evaluate and compare the robustness against non-constancy of different designs and success criteria (see Table~\ref{tab:l0min}). The traditional synthesis method has $\lambda_{0,\text{min}} = 0$, ensuring protected type-I error under constancy but inflated error when $\lambda_0<0$. The bias-adjusted synthesis method has $\lambda_{0,\text{min}}=\lambda_1$ and the Odem-Davis method has $\lambda_{0,\text{min}}$ strictly smaller than $\lambda_1$. Hence, both methods have protected (conservative) type-I error when the non-constancy level is less extreme than the assumed non-constancy level $\lambda_1$. Notably, $\lambda_1<0$ means that the active control in the target population is assumed to be less effective than in the historical population. The 95-95 method has $\lambda_{0,\text{min}}$ strictly smaller than 0. Hence, it has protected type-I error under constancy, and allows for some degree of non-constancy. The amount of non-constancy accommodated depends on the uncertainty in $\hat\gamma_{CP,H}$ as shown in Appendix~\ref{sec:appA_tol_non_constancy}. The 0-95 method has $\lambda_{0,\text{min}}>0$, making it anti-conservative under constancy.

The tolerable levels of non-constancy for the different methods examined in our HIV prevention trial design example are provided in the Controlled non-constancy column of Tables~\ref{tab:comp_methods_preserv} and~\ref{tab:comp_methods_inf_eff}, expressed on the prevention efficacy scale. These values represent the lowest true prevention efficacy of CAB-LA in the target population for which unconditional type I error is controlled when rejection is based on the success margin.

\subsection{Choosing Success Criteria}\label{sec:comp_criteria}

A final insight emerges from the general framework. While the formulas~\eqref{up_um}--\eqref{eq:ct1e_um} enable numerical comparisons of operating characteristics across different success criteria, it is essential to ensure that their scientific null hypotheses are aligned numerically for a fair comparison in terms of type-I error. Without such alignment, the type-I error rates are not comparable, as they correspond to different scientific null hypotheses.

For the preservation of effect criterion, the scientific null hypothesis depends on the true active control effect within the target population. Thus, even if a preservation of effect criterion and an inferred efficacy criterion are aligned so that they correspond to the same scientific null hypothesis, this alignment only holds for a particular value of the active control effect. Consequently, we advocate for selecting the success criterion based on scientific rationale rather than solely based on a comparison of operating characteristics.

In our HIV prevention trial design example, this consideration informed our use of two distinct criteria. Preserving 50\% of a 92.8\% CAB-LA PE on the log scale corresponds to a 73.2\% inferred PE, while requiring an inferred PE of 30\% preserves only 13.6\% of the CAB-LA effect. These targets are not numerically aligned, but they reflect different scientific goals and policy questions---one prioritizing relative preservation of an established intervention’s efficacy, the other focusing on achieving a minimum level of absolute protection with meaningful public health impact.

\subsection{Systematic Comparison within the General Framework}

We now walk through the trial design results in Tables~\ref{tab:comp_methods_preserv} and~\ref{tab:comp_methods_inf_eff}, highlighting important comparisons. For example, using the 95-95 method with the $50\%$ preservation of effect criterion, achieving $90\%$ unconditional power requires 15,516 participants (7,758 per arm) to accrue 52 events. The experimental intervention is deemed successful if the hazard of HIV acquisition in this group is up to 2.05 times that in the control group. This design controls unconditional type-I error below 0.025 as long as CAB-LA PE is at least $85.9\%$ ($\lambda_{0,\min}=-0.26$), and it reaches $70\%$ unconditional power when the true CAB-LA PE is higher than expected, at $94.7\%$. In contrast, the traditional approach requires 11,012 but achieves only $83\%$ unconditional power under constancy and $60\%$ unconditional power if CAB-LA PE is $94.7\%$.
 
 Evaluating these trial designs shows that sizing a trial to achieve $90\%$ conditional power yields unconditional power below $90\%$, confirming the risk of underpowered trials. The ad hoc approach, targeting $90\%$ conditional power under a higher CAB-LA PE of $94.7\%$, generally requires sample sizes at least 1.5 times larger than the traditional approach based on conditional power, often achieving over $90\%$ unconditional power but at a higher resource cost. The novel approach controlling unconditional power strikes a balance between the two aforementioned approaches. Additionally, the unconditional approach demonstrates greater robustness to non-constancy in terms of unconditional power compared to the traditional approach. Unconditional power consistently exceeds 70$\%$ using the preservation of effect criterion, and exceeds 75$\%$ with the inferred efficacy criterion, when CAB-LA PE is 94.7$\%$. This underscores the utility of the novel approach in ensuring adequate power while maintaining manageable sample sizes and ensuring robustness to non-constancy across various scenarios.

When comparing the five methods within the same success criterion and design approach, the 0–95 method yields the most lenient margin and the smallest sample size, while also demonstrating the greatest robustness to non-constancy in terms of power loss. However, as anticipated, it inflates type I error even under constancy, since it has a tolerable level of non-constancy $\lambda_{0,\min}>0$. In contrast, the Odem–Davis method accommodates the highest level of non-constancy, and in most scenarios achieves higher power than the 95–95 method. The bias-adjusted synthesis method generally maintains good robustness to non-constancy in terms of power, while controlling type I error for the assumed relative effect deviation, since its tolerable level of non-constancy is $\lambda_{0,\min} = \lambda_1$.

 Lastly, for a given method and design approach, the $50\%$ preservation of effect criterion sets stricter margins and typically requires larger sample sizes compared to the inferred $30\%$ efficacy criterion,  except with the Odem-Davis method under unconditional power control.

Overall, this application shows that: 1) with reasonable assumptions, the bias-adjusted synthesis method is efficient and accommodates substantial non-constancy, 2) the 95-95 method does not generally outperform others in robustness to non-constancy, 3) the inferred efficacy non-inferiority criterion is attractive for its interpretability and practicality, and 4) the approach that controls unconditional power effectively addresses uncertainty in the active control effect.

\section{Discussion}

In this article, we present a general framework for the design of active-controlled trials evaluating the non-inferiority of an experimental intervention that encompasses a wide variety of methods and success criteria, and accommodates binary, continuous, count, and censored event-time outcomes. A key contribution of our framework is facilitating rational decision-making: it allows researchers to select the most suitable design based on the trade-off between type-I error and power and to evaluate the level of non-constancy that is accommodated by the design. The framework includes the novel inferred efficacy non-inferiority criterion as a special case. Additionally, it introduces a novel approach to trial design, prioritizing the control of unconditional power to ensure robustness to non-constancy.

The development of a framework for designing non-inferiority trials to evaluate the inferred efficacy criterion will advance future trial design. In particular, when an active control is highly efficacious, and an experimental intervention offers advantages in terms of acceptability or feasibility, a traditional preservation of effect margin may be overly stringent. The inferred efficacy criterion is also highly interpretable.

Our framework also clarifies and generalizes previous conceptual unifications of non-inferiority procedures, such as Snapinn's discounting perspective. Snapinn interpreted preservation of effect criteria and fixed-margin methods as ways to down-weight historical evidence to protect against untestable assumptions (e.g., constancy), with the ultimate goal of demonstrating superiority relative to a hypothetical placebo. In contrast, our framework explicitly parameterizes the assumed degree of non-constancy and how the uncertainty from the historical data is incorporated into the analysis, making the underlying assumptions explicit. This allows researchers to distinguish genuine preservation of effect or inferred efficacy objectives from methods that merely discount historical data. Our framework also emphasizes that the common 95-95 approach assumes a varying degree of non-constancy, depending on the precision of the historical active control effect estimate. This may be unappealing in situations in which the historical estimate is precise but unlikely to bridge to the target population because of variation in relevant effect modifiers. Consequently, our framework provides a clearer, quantitative foundation for comparing existing methods and designing robust non-inferiority trials.

While our framework encompasses both the traditional preservation of effect criterion and the innovative inferred efficacy criterion, caution is warranted when directly comparing these criteria, as they typically correspond to different scientific null hypotheses. This distinction is important because previous studies suggest that preservation of effect criteria may demonstrate greater robustness to non-constancy violations than the inferred efficacy criterion \citep{wang2002utility}. However, our work suggests that this is not true in general.  Moreover, even if attempts are made to align the scientific null hypotheses, as discussed in Section~\ref{sec:comp_criteria}, this alignment is only valid for a specific value of the active control effect. This underscores the need to carefully consider the nuances of each criterion and their associated scientific null hypotheses when designing the trial and interpreting its outcomes.

Importantly, we considered the setting in which an active control is effective and available to the population of interest.  However, in settings where the active control is not available or not acceptable to the population, a placebo-controlled design remains the gold standard for generating evidence regarding efficacy of the experimental intervention.

Overall, our framework advances the understanding and application of statistical methods for non-inferiority assessment, offering researchers a powerful toolkit for rigorous non-inferiority study design.

\section*{Acknowledgments}
This work was supported by the National Institute of Allergy and Infectious Diseases grant UM1AI068635.

\bibliographystyle{apalike}
\bibliography{references}

\appendix

\section{Derivation of the Formulas for computing the Operating Characteristics}
\vspace*{12pt}

\subsection{Formulas for Unconditional and Conditional Power}\label{sec:appA_formulas}
\vspace*{12pt}

We begin by deriving the distribution of the test statistic under the assumed model. Since $\hat\gamma_{CP,H}\sim \mathcal{N}\left(\gamma_{CP,H},V_{CP,H}\right)$ and $\hat\gamma_{XC}\sim \mathcal{N}\left(\gamma_{XC},V_{XC}\right)$, the unconditional distribution of $\hat\gamma_{XC}+(1-f)(1+\lambda_1)\hat\gamma_{CP,H}$ is
\[
\mathcal{N}\left(\gamma_{XC}+(1-f)(1+\lambda_1)\gamma_{CP,H},V_{XC}+(1-f)^2\widetilde{V}_{CP,H}\right),
\]
where
\[
\widetilde{V}_{CP,H}=\begin{cases}
    (1+\lambda_1)^2V_{CP,H},& \text{if }u>0,  \\[6pt]
    V_{CP,H}, & \text{if } u=0 \text{ (fixed-margin methods).} 
\end{cases}
\]
These two cases for $\widetilde{V}_{CP,H}$ reflect whether the margin component contributes randomness: in fixed-margin methods ($u=0$), the term $\lambda_1\hat\gamma_{CP,H}=Z_{1-\theta}\sqrt{V_{CP,H}}$ is deterministic and does not add variance, whereas for $u>0$ the full term $(1+\lambda_1)\hat\gamma_{CP,H}$ is random, inflating the variance by $(1+\lambda_1)^2$. Writing the variance contribution in terms of $\widetilde{V}_{CP,H}$ allows both situations to be handled with a single expression.

Hence, the unconditional distribution of $T_{u,\lambda_1,f,\Delta_0}$ is
\[
\mathcal{N}\left(\frac{\gamma_{XC}+(1-f)(1+\lambda_1)\gamma_{CP,H}-\Delta_0}{\sqrt{V_{XC}+u^2(1-f)^2(1+\lambda_1)^2V_{CP,H}}}, \sigma^2\right),
\]
 where $\sigma^2=\big\{V_{XC}+(1-f)^2\widetilde{V}_{CP,H}\big\}/\left\{V_{XC}+u^2(1-f)^2(1+\lambda_1)^2V_{CP,H}\right\}$. Moreover, at the boundary of the operational null hypothesis (\ref{h0_op}), the distribution of $T_{u,\lambda_1,f,\Delta_0}$ has mean zero.

Next, the unconditional power of  $T_{u,\lambda_1,f,\Delta_0}$ to reject the scientific null hypothesis (\ref{h0}) when the experimental intervention has effect size $\gamma_{XP}$ is
\begin{align*}
    \text{Pr}&\left( T_{u,\lambda_1,f,\Delta_0} <-Z_{1-\alpha}\right)\\
    &=\text{Pr}\left\{\frac{\hat\gamma_{XC}+(1-f)(1+\lambda_1)\hat\gamma_{CP,H}-\Delta_0}{\sqrt{V_{XC}+u^2(1-f)^2(1+\lambda_1)^2V_{CP,H}}}<-Z_{1-\alpha}\right\}\\
     &=\text{Pr}\left\{\hat\gamma_{XC}+(1-f)(1+\lambda_1)\hat\gamma_{CP,H}<\Delta_0-Z_{1-\alpha}\sqrt{V_{XC}+u^2(1-f)^2(1+\lambda_1)^2V_{CP,H}}\right\}\\
     &\stackrel{(\text{i})}{=}\text{Pr}\left\{Z<\frac{\Delta_0-(1-f)(1+\lambda_1)\gamma_{CP,H}-\gamma_{XC}-Z_{1-\alpha}\sqrt{V_{XC}+u^2(1-f)^2(1+\lambda_1)^2V_{CP,H}}}{\sqrt{V_{XC}+(1-f)^2\widetilde{V}_{CP,H}}}\right\}\\
    &=\Phi\left(\frac{\Delta_0-(1-f)(1+\lambda_1)\gamma_{CP,H}-\gamma_{XC}-Z_{1-\alpha}\sqrt{V_{XC}+u^2(1-f)^2(1+\lambda_1)^2V_{CP,H}}}{\sqrt{V_{XC}+(1-f)^2\widetilde{V}_{CP,H}}}\right)\\
     &\stackrel{(\text{ii})}{=}\Phi\left(\frac{\Delta_0+\{(1+\lambda_0)-(1-f)(1+\lambda_1)\}\gamma_{CP,H}-\gamma_{XP}-Z_{1-\alpha}\sqrt{V_{XC}+u^2(1-f)^2(1+\lambda_1)^2V_{CP,H}}}{\sqrt{V_{XC}+(1-f)^2\widetilde{V}_{CP,H}}}\right),
\end{align*}
where in the step (i) we first subtracted $\gamma_{XC}+(1-f)(1+\lambda_1)\gamma_{CP,H}$ and then divide by $\sqrt{V_{XC}+(1-f)^2\widetilde{V}_{CP,H}}$ in both sides, and in (ii) we used that $\gamma_{XC}=\gamma_{XP}-\gamma_{CP}=\gamma_{XP}-(1+\lambda_0)\gamma_{CP,H}$.

On the other hand, assuming $\gamma_{CP,H}=\hat\gamma_{CP,H}$ and following analogous steps, the conditional power of $T_{u,\lambda_1,f,\Delta_0}$ to reject the scientific null hypothesis (\ref{h0}) when the experimental intervention has effect size $\gamma_{XP}$ is
\begin{align*}
    \text{Pr}&\left( T_{u,\lambda_1,f,\Delta_0} <-Z_{1-\alpha}\right) \\
    &=\Phi\left(\frac{\Delta_0+\{(1+\lambda_0)-(1-f)(1+\lambda_1)\}\hat\gamma_{CP,H}-\gamma_{XP}-Z_{1-\alpha}\sqrt{V_{XC}+u^2(1-f)^2(1+\lambda_1)^2V_{CP,H}}}{\sqrt{V_{XC}}}\right),
\end{align*}
where the assumption that $\gamma_{CP,H}=\hat\gamma_{CP,H}$ is justified because $\hat\gamma_{CP,H}$ is treated as fixed, with mean equal to $\gamma_{CP,H}$.

For fixed-margin methods, where $u=0$ and $\lambda_1=Z_{1-\theta}\sqrt{V_{CP,H}}/\hat\gamma_{CP,H}$, the previous expression reduces to
\begin{align*}
\Phi\left(\frac{\Delta_0-(1-f)(1+\lambda_1)\hat\gamma_{CP,H}-\gamma_{XC}-Z_{1-\alpha}\sqrt{V_{XC}}}{\sqrt{V_{XC}}}\right)=\Phi\left(-Z_{1-\alpha}+\frac{\delta-\gamma_{XC}}{\sqrt{V_{XC}}}\right),
\end{align*}
where $\delta$ is the success margin defined in (\ref{ni_margin}). This expression is commonly used to compute the required precision, $V_{XC}$, of the relative effect of experimental versus active control intervention that ensures adequate conditional power and type I error control under constancy.

Finally, type I error probability is obtained by evaluating the power expressions under the null boundary values. Specifically, when $\gamma_{XP}=\Delta_0+f\:(1+\lambda_0)\gamma_{CP,H}$, the expressions for unconditional and conditional power reduce to expressions~\eqref{eq:ut1e_um} and~\eqref{eq:ct1e_um}, respectively.

\subsection{Detectable Design Alternatives and Maximum Unconditional Power}\label{sec:appA_parameter_space}
\vspace*{12pt}

For a given design alternative $\gamma_{XP}$ to be detectable with power exceeding 50\%, the numerator inside the function $\Phi(\cdot)$ in expressions~\eqref{up_um} and~\eqref{cp_um} must be positive. Below, we provide necessary and sufficient conditions for this to occur, and we clarify the different implications for conditional versus unconditional power.

Define $x:=\sqrt{V_{XC}}$, $a:=\Delta_0+\left\{(1+\lambda_0)-(1-f)(1+\lambda_1)\right\}\gamma_{CP,H}-\gamma_{XP}$, and $\varepsilon:=(1-f)(1+\lambda_1)\sqrt{V_{CP,H}}$. For any $\varepsilon>0$, $x>0$, and $u\ge0$, it holds that $u\varepsilon < \sqrt{x^2+u^2\varepsilon^2}$, so
\[
a-Z_{1-\alpha}\sqrt{x^2+u^2\varepsilon^2}<a-u\varepsilon Z_{1-\alpha}\quad \forall x>0.
\]
Hence, if $a-u\varepsilon Z_{1-\alpha}\le0$, then $a-Z_{1-\alpha}\sqrt{x^2+u^2\varepsilon^2}< 0$ for all $x>0$, implying that both conditional and unconditional powers are less than 0.5.

Conversely, for any $u\ge 0$ and $x>0$, we have
\begin{align*}
    \frac{a-Z_{1-\alpha}\sqrt{x^2+u^2\varepsilon^2}}{x}&=\frac{a}{x}-Z_{1-\alpha}\sqrt{1+\frac{u^2\varepsilon^2}{x^2}}\\
    &\ge \frac{a}{x}-Z_{1-\alpha}\left(1+\frac{u\varepsilon}{x}\right)\\
    &=-Z_{1-\alpha}+\frac{a-u\varepsilon Z_{1-\alpha}}{x}.
\end{align*}
If $a-u\varepsilon Z_{1-\alpha}>0$, the last expression can be made arbitrarily positive by choosing $x$ appropriately.  Consequently, the conditional power can be made arbitrarily close to 1. Thus, the inequality $a-u\varepsilon Z_{1-\alpha}>0$ is both necessary and sufficient for conditional power to exceed 50\%. In terms of the original parameters, the condition is
\[
\gamma_{XP}<\Delta_0+(1+\lambda_0)\gamma_{CP,H}-(1-f)(1+\lambda_1)\left(\gamma_{CP,H}+uZ_{1-\alpha}\sqrt{V_{CP,H}}\right).
\]

 The situation is more restrictive for unconditional power. Define $\tilde\varepsilon:=(1-f)\sqrt{\widetilde{V}_{CP,H}}$. For any $x>0$, we have $\tilde\varepsilon < \sqrt{x^2+\tilde\varepsilon^2}$, which implies
\[
\frac{ a-Z_{1-\alpha}\sqrt{x^2+u^2\varepsilon^2}}{\sqrt{x^2+\tilde\varepsilon^2}}<\frac{a-u\varepsilon Z_{1-\alpha}}{\tilde\varepsilon}=\frac{a}{\tilde\varepsilon}-uZ_{1-\alpha},
\]
where the equality in the previous display holds because $\varepsilon=\tilde\varepsilon$ when $u>0$. Thus, unconditional power is bounded above by
\[
        \Phi\left(\frac{a}{\tilde\varepsilon}-uZ_{1-\alpha}\right)=\Phi\left(-uZ_{1-\alpha }+\frac{\Delta_0+\left\{(1+\lambda_0)-(1-f)(1+\lambda_1)\right\}\gamma_{CP,H}-\gamma_{XP}}{(1-f)\sqrt{\widetilde{V}_{CP,H}}}\right).
       \]
       Unlike conditional power, unconditional power cannot be increased without limit by adjusting $V_{XC}$; instead, its maximum value is determined by the bound above.
       
Finally, to detect a design alternative with an unconditional power of at least $1-\beta$, the following condition must hold:
\[
\frac{a}{\tilde \varepsilon}-uZ_{1-\alpha}>Z_{1-\beta},
\]
which is equivalent to
\[
\gamma_{XP}<\Delta_0+(1+\lambda_0)\gamma_{CP,H}-(1-f)\left\{(1+\lambda_1)\gamma_{CP,H}+(uZ_{1-\alpha}+Z_{1-\beta})\sqrt{\widetilde{V}_{CP,H}}\right\}.
\]
This condition is necessary and sufficient for achieving unconditional power $1-\beta$.

\subsection{Acceptable Assumed Level of Non-Constancy and Tolerable Level of Non-Constancy}\label{sec:appA_tol_non_constancy}
\vspace*{12pt}

In this subsection, we (i) derive conditions on the assumed relative effect deviation \(\lambda_1\) that guarantee the unconditional type I error remains below 0.5 for all \(V_{XC}>0\), and (ii) determine the tolerable level of non-constancy \(\lambda_{0,\min}\) that a given method (specified by \(u\) and \(\lambda_1\)) can accommodate while controlling unconditional type I error at the nominal level.

To ensure that the unconditional type I error remains below 0.5, the numerator in expression~\eqref{eq:ut1e_um} must be negative. This yields the requirement

\[
(1-f)(\lambda_0-\lambda_1)\gamma_{CP,H}<Z_{1-\alpha}\sqrt{V_{XC}+u^2(1-f)^2(1+\lambda_1)^2V_{CP,H}}.
\]
Since $V_{XC}$ can take any positive value, we seek conditions under which this inequality holds for all $V_{XC}>0$. This gives
\begin{equation}
    (1-f)(\lambda_0-\lambda_1)\gamma_{CP,H}\le u(1-f)(1+\lambda_1)Z_{1-\alpha}\sqrt{V_{CP,H}}.\label{eq:ineq_cnc}
\end{equation}

If $\lambda_1\le\lambda_0$, the inequality~\eqref{eq:ineq_cnc} always holds, given that $\gamma_{CP,H}$ is assumed negative. Conversely, when $\lambda_1>\lambda_0$, and assuming $\gamma_{CP,H}+uZ_{1-\alpha}\sqrt{V_{CP,H}}<0$, the inequality \eqref{eq:ineq_cnc} is satisfied if
\[
\lambda_1\le\lambda_0-\frac{(1+\lambda_0)uZ_{1-\alpha}\sqrt{V_{CP,H}}}{\gamma_{CP,H}+uZ_{1-\alpha}\sqrt{V_{CP,H}}}.
\]
The assumption $\gamma_{CP,H}+uZ_{1-\alpha}\sqrt{V_{CP,H}}<0$ is justified because the active control is effective in the historical setting. Specifically, if $\gamma_{CP,H}=\hat\gamma_{CP,H}$, then this expression represents the upper bound of a confidence interval for $\gamma_{CP,H}$, which we expect to lie below zero.

To quantify the level of non-constancy that a given method in the general framework can properly accommodate, we solve
\[
\frac{(1-f)(\lambda_0-\lambda_1)\gamma_{CP,H}-Z_{1-\alpha}\sqrt{V_{XC}+u^2(1-f)^2(1+\lambda_1)^2V_{CP,H}}}{\sqrt{V_{XC}+(1-f)^2\widetilde{V}_{CP,H}}}\le -Z_{1-\alpha}
\]
for $\lambda_0$. Given that $\gamma_{CP,H}<0$, the solution is $\lambda_0\ge \lambda_{0,\min}$, where
\begin{equation*}
    \lambda_{0,\min} := \lambda_1+\frac{\sqrt{V_{XC}+u^2(1-f)^2(1+\lambda_1)^2V_{CP,H}}-\sqrt{V_{XC}+(1-f)^2\widetilde{V}_{CP,H}}}{(1-f)\gamma_{CP,H}}Z_{1-\alpha}.
\end{equation*}
This value, $\lambda_{0,\min}$, defines the tolerable level of non-constancy for the method.

For the bias-adjusted synthesis method $(u=1)$, $\lambda_{0,\min}=\lambda_1$. For the Odem-Davis method ($u=\{1+\lambda_1\}^{-1}$), we have
\begin{align*}
    \sqrt{V_{XC}+u^2(1-f)^2(1+\lambda_1)^2V_{CP,H}}&=\sqrt{V_{XC}+(1-f)^2V_{CP,H}}\\
    &>\sqrt{V_{XC}+(1-f)^2(1+\lambda_1)^2V_{CP,H}}\\
    &=\sqrt{V_{XC}+(1-f)^2\widetilde{V}_{CP,H}}
\end{align*}
for any $\lambda_1\in(-1,0)$. Since $\gamma_{CP,H}<0$, it follows that $\lambda_{0,\min}<\lambda_1$.

For fixed-margin methods, where $u=0$ and $\lambda_1=Z_{1-\theta}\sqrt{V_{CP,H}}/\hat\gamma_{CP,H}$, we obtain
\[
  \lambda_{0,\min} =\frac{Z_{1-\theta}\sqrt{V_{CP,H}}}{\hat\gamma_{CP,H}}+ \frac{\sqrt{V_{XC}}-\sqrt{V_{XC}+(1-f)^2V_{CP,H}}}{(1-f)\gamma_{CP,H}}Z_{1-\alpha}
\]
since $\widetilde{V}_{CP,H}=V_{CP,H}$ when $u=0$. Because $\gamma_{CP,H}$ is assumed equal to $\hat\gamma_{CP,H}$ for fixed-margin methods, this can be rewritten as
\begin{equation}
\lambda_{0,\min} = \frac{\sqrt{V_{XC}}+(1-f)\frac{Z_{1-\theta}}{Z_{1-\alpha}}\sqrt{V_{CP,H}}-\sqrt{V_{XC}+(1-f)^2V_{CP,H}}}{(1-f)\gamma_{CP,H}}Z_{1-\alpha}
   .\label{l0_fm}
\end{equation}
Finally, since $\gamma_{CP,H}<0$, the right-hand side of~\eqref{l0_fm} is negative for the 95-95 method ($\theta=\alpha=0.025$), indicating that $\lambda_{0,\min}<0$. In contrast, for the 0-95 method ($Z_{1-\theta}=0$), the right-hand side of~\eqref{l0_fm} becomes positive, resulting in $\lambda_{0,\min}>0$.

\section{Vignette Application to Designing an Active-Controlled Trial}
\vspace*{12pt}\label{sec:r_function}

We provide R code in the GitHub repository \href{https://github.com/aolivasm/ni-design}{\texttt{github.com/aolivasm/ni-design}} demonstrating the use of our \texttt{ni\_design()} R function for generating non-inferiority trial design specifications and computing operating characteristics under the preservation of effect and inferred efficacy criteria when the effect is measured in the log hazards ratio scale. The function computes non-inferiority margins, required numbers of events, sample sizes, and unconditional power for a set of analytical methods defined by $(u,\lambda_1)$ pairs.

The \texttt{ni\_design()} function requires the following inputs:
\begin{itemize}
\item \texttt{u.list}: Vector of $u$ parameters specifying the desired analytical methods.
\item \texttt{l1.list}: Vector of $\lambda_1$ parameters specifying the desired analytical methods. Must have the same length as \texttt{u.list}.
Examples include: $(u,\lambda_1)=(1,0)$ for the traditional synthesis method, and $(u,\lambda_1)=(0,1.96\sqrt{V_{CP,H}}/\hat\gamma_{CP,H})$ for the 95–95 method.
\item \texttt{design.alternative.pe}: True prevention efficacy (PE) of the experimental treatment relative to placebo under the design alternative. Must be specified on the PE scale.
\item \texttt{hist.ac.pe}: Historical estimate of active control prevention efficacy (PE scale).
\item \texttt{hist.ac.effect.se}: Standard error of the historical active control effect estimate (log hazard ratio scale).
\end{itemize}

Optional inputs and defaults
\begin{itemize}
\item \texttt{f.preserv}: Fraction of active control effect to preserve for the preservation-of-effect criterion (default: \texttt{0.5}).
\item \texttt{null.pe}: Null prevention efficacy for the inferred-efficacy criterion (default: \texttt{0.3}).
\item \texttt{lambda0.for.design}: Value of $\lambda_0$	 (relative effect deviation) assumed for the primary design scenario (default: \texttt{0}).
\item \texttt{target.on.unconditional.power}: Logical flag; if \texttt{TRUE}, designs target unconditional power (default: \texttt{TRUE}).
\item \texttt{allocation.ratio}: Experimental:control allocation ratio (default: \texttt{1}).
\item \texttt{power}: Desired power (default: \texttt{0.9}).
\item \texttt{sign.level}: One-sided significance level (default: \texttt{0.025}).
\item \texttt{lambda0.sens.analysis}: Optional $\lambda_0$	
  for sensitivity analysis of unconditional power.
\item \texttt{placebo.incidence.rate}: Annual incidence rate in the placebo population (default: \texttt{0.03}).
\item \texttt{loss.to.followup}: Annual loss-to-follow-up proportion (default: \texttt{0.075}).
\item \texttt{trial.duration}: Planned trial duration in years (default: \texttt{2}).
\item \texttt{correction}: Logical flag; if \texttt{TRUE}, applies correction for interim monitoring (default: \texttt{FALSE}).
\end{itemize}

The function returns an object of class \texttt{ni.design}, which is a list containing:
\begin{itemize}
\item \textbf{Specifications}: A character vector summarizing the design approach and, if applicable, the sensitivity analysis assumptions.
\item \textbf{Data frames for each success criterion}: For example, one table for preservation-of-effect and one for inferred efficacy. Each table contains:
\begin{itemize}
\item \texttt{Method}: Name of the analytical method.
\item \texttt{NI margin}: Computed non-inferiority margin.
\item \texttt{RNE}, \texttt{Exp}, \texttt{Ctr}: Required number of events (total, experimental, and control arms).
\item \texttt{Sample size}, \texttt{Exp.arm}, \texttt{Ctr.arm}: Required sample sizes.
\item \texttt{CNC}: Control non-constancy—the minimum active control PE for which type-I error is controlled.
\item \texttt{U.power}: Unconditional power under the design scenario.
\item \texttt{U.power (SA)}: Unconditional power under the sensitivity-analysis scenario (if specified).
\end{itemize}
\end{itemize}

The custom \texttt{summary()} function displays the trial specifications followed by the design result tables for each non-inferiority criterion.

\subsection{Example Usage}
\vspace*{12pt}
The following code illustrates the design considered in this manuscript using the traditional approach that targets conditional power:

\begin{verbatim}
    obj.designs.app1 = ni_design(u.list = c(1, 1, 1/(1-0.23), 0, 0),
                              l1.list = c(0, -0.23, -0.23, 1.96*0.61/log(1-0.928), 0),
                              f.preserv = 0.5,
                              null.pe = 0.3,
                              design.alternative.pe = 0.95,
                              hist.ac.pe = 0.928,
                              hist.ac.effect.se = 0.61,
                              lambda0.for.design = 0,
                              target.on.unconditional.power = FALSE,
                              allocation.ratio = 1,
                              power = 0.9, sign.level = 0.025,
                              lambda0.sens.analysis = 0.12,
                              placebo.incidence.rate = 0.03,
                              loss.to.followup = 0.075,
                              trial.duration = 2,
                              correction = FALSE)

\end{verbatim}

The output using the \texttt{summary()} function is:

{\fontsize{7.5pt}{11pt}\selectfont
\begin{verbatim}
    summary(obj.designs.app1)

    === Summary of Non-Inferiority Trial Design ===

Design Specifications:
 • Approach : Design approach targeting 90% conditional power and assuming an active control efficacy of 92.8% 
 • Sensitivity analysis : Sensitivity analysis (SA) assumes an active control efficacy of 94.7% 

--- NI criterion: Preserving 50% of active control effect ---


|     Method     | NI margin | RNE | Exp | Ctr | Sample size | Exp.arm | Ctr.arm |  CNC  | U.power | U.power (SA) |
|:--------------:|:---------:|:---:|:---:|:---:|:-----------:|:-------:|:-------:|:-----:|:-------:|:------------:|
| Traditional SM |   3.12    | 19  |  8  | 11  |    5766     |  2883   |  2883   | 0.928 |  0.86   |     0.69     |
| BA-SM, lm1=-23% |   2.42    | 27  | 11  | 16  |    8008     |  4004   |  4004   | 0.868 |  0.86   |     0.65     |
|  OD, lm1=-23%   |   2.21    | 32  | 13  | 19  |    9510     |  4755   |  4755   | 0.844 |  0.86   |     0.63     |
|  95-95 method  |   2.05    | 37  | 15  | 22  |    11012    |  5506   |  5506   | 0.850 |  0.83   |     0.60     |
|  0-95 method   |   3.73    | 15  |  6  |  9  |    4504     |  2252   |  2252   | 0.948 |  0.87   |     0.72     |

--- NI criterion: Inferred efficacy of 30% relative to hypothetical placebo ---


|     Method     | NI margin | RNE | Exp | Ctr | Sample size | Exp.arm | Ctr.arm |  CNC  | U.power | U.power (SA) |
|:--------------:|:---------:|:---:|:---:|:---:|:-----------:|:-------:|:-------:|:-----:|:-------:|:------------:|
| Traditional SM |   6.13    |  9  |  4  |  5  |    2882     |  1441   |  1441   | 0.928 |  0.83   |     0.73     |
| BA-SM, lm1=-23% |   3.72    | 15  |  6  |  9  |    4504     |  2252   |  2252   | 0.868 |  0.83   |     0.69     |
|  OD, lm1=-23%   |   2.88    | 22  |  9  | 13  |    6506     |  3253   |  3253   | 0.837 |  0.81   |     0.65     |
|  95-95 method  |   2.94    | 21  |  9  | 12  |    6486     |  3243   |  3243   | 0.870 |  0.78   |     0.63     |
|  0-95 method   |   9.72    |  7  |  3  |  4  |    2162     |  1081   |  1081   | 0.952 |  0.85   |     0.76     |
\end{verbatim}
}
\nocite{*}

\end{document}